\documentclass[12pt]{iopart}
\usepackage{iopams}
\usepackage{amsmath,amssymb,color,comment,physics}
\usepackage[makeroom]{cancel}
\usepackage[caption=false]{subfig}
\usepackage{mathrsfs}
\usepackage{graphicx}
\usepackage{subfig}
\usepackage[countmax]{subfloat}
\usepackage[english]{babel}
\usepackage[bookmarks=true,colorlinks,linkcolor=OrangeRed,urlcolor=NavyBlue,citecolor=RoyalBlue]{hyperref}
\usepackage[dvipsnames]{xcolor}
\usepackage{ulem} 
 \usepackage{dsfont}

\usepackage{etoolbox}
\usepackage{hyperref}

\usepackage{lipsum}

\makeatletter
\def\@mkboth#1#2{}
\newlength\appendixwidth
\preto\appendix{\addtocontents{toc}{\protect\patchl@section}}
\newcommand{\patchl@section}{%
	\settowidth{\appendixwidth}{\textbf{Appendix }}%
	\addtolength{\appendixwidth}{1.5em}%
	\patchcmd{\l@section}{1.5em}{\appendixwidth}{}{\ddt}%
}
\makeatother
\definecolor{mygreen}{rgb}{0,0.5,0}
\definecolor{myblue}{rgb}{0,0,0.75}
\definecolor{mymagenta}{cmyk}{0,1,0,0.12}
\definecolor{mygray}{rgb}{0.5,0.5,0.5}

\definecolor{mypink1}{rgb}{0.858, 0.188, 0.478}
\definecolor{mypurple}{rgb}{0.49,0.18,0.56}
\definecolor{mygold}{rgb}{0.93,0.69,0.13}
\definecolor{mygreen}{rgb}{0,0.5,0}
\definecolor{myblue}{rgb}{0,0,0.75}
\definecolor{mymagenta}{cmyk}{0,1,0,0.12}
\definecolor{mygray}{rgb}{0.5,0.5,0.5}

\newcommand{\canc}[1]{}

\newcommand{\ignore}[1]{}
\usepackage{geometry}

\begin{document}
\title{Gauge protection in non-Abelian lattice gauge theories}
\author{Jad C.~Halimeh}
\address{INO-CNR BEC Center and Department of Physics, University of Trento, Via Sommarive 14, I-38123 Trento, Italy}

\author{Haifeng Lang}
\address{INO-CNR BEC Center and Department of Physics, University of Trento, Via Sommarive 14, I-38123 Trento, Italy}
\address{Theoretical Chemistry, Institute of Physical Chemistry,
Heidelberg University, Im Neuenheimer Feld 229, 69120 Heidelberg, Germany}

\author{Philipp Hauke}
\address{INO-CNR BEC Center and Department of Physics, University of Trento, Via Sommarive 14, I-38123 Trento, Italy}

\begin{abstract}
Protection of gauge invariance in experimental realizations of lattice gauge theories based on energy-penalty schemes has recently stimulated impressive efforts both theoretically and in setups of quantum synthetic matter. A major challenge is the reliability of such schemes in non-Abelian gauge theories where local conservation laws do not commute. Here, we show through exact diagonalization that non-Abelian gauge invariance can be reliably controlled using gauge-protection terms that energetically stabilize the target gauge sector in Hilbert space, suppressing gauge violations due to unitary gauge-breaking errors. We present analytic arguments that predict a volume-independent protection strength $V$, which when sufficiently large leads to the emergence of an \textit{adjusted} gauge theory with the same local gauge symmetry up to least a timescale $\propto\sqrt{V/V_0^3}$. Thereafter, a \textit{renormalized} gauge theory dominates up to a timescale $\propto\exp(V/V_0)/V_0$ with $V_0$ a volume-independent energy factor, similar to the case of faulty Abelian gauge theories. Moreover, we show for certain experimentally relevant errors that single-body protection terms robustly suppress gauge violations up to all accessible evolution times in exact diagonalization, and demonstrate that the adjusted gauge theory emerges in this case as well. These single-body protection terms can be readily implemented with fewer engineering requirements than the ideal gauge theory itself in current ultracold-atom setups and NISQ devices.
\end{abstract}

\date{\today}
\maketitle
\makeatletter
\tableofcontents

\section{Introduction}
In the 1980’s, Feynman proposed a possible solution to computationally hard quantum many-body problems, namely to engineer the abstract model in a designed, well-controlled quantum machine and thus compute the model dynamics by measurement in a laboratory device \cite{Feynman1982,Lloyd1996}. Meanwhile, this vision---now known under the name of quantum simulation---has become a reality, and many implementations exist especially in quantum systems whose elementary units consist of ultracold atoms \cite{Bloch_review,Schaefer2020}, trapped ions \cite{Schneider2012,Blatt_review,Monroe2021}, superconducting qubits \cite{Houck_review,Kjaergaard2020}, or photons \cite{Aspuru-Guzik_review,Ozawa2019}. Initially, the focus of the community lay on models from condensed matter physics \cite{Lewenstein_review,Hauke2012,Cirac2012,Georgescu2014}.
Lately, however, another type of theories has attracted the attention of researchers: gauge theories \cite{Wiese_review,Zohar2015,Dalmonte2016,MariCarmen2019,Alexeev2021,Zohar2021quantum}.

These theories describe the interactions of elementary particles and are thus of large importance for our fundamental description of nature \cite{Weinberg_book,Gattringer_book,Cheng_book,Zee_book}, and they find relevance as emerging theories at low temperatures \cite{Hastings2005,Sachdev2018}. Yet, their dynamics is extremely difficult to solve with numerically unbiased methods, especially at strong coupling. These features make gauge theories particularly rewarding targets for quantum simulation. Even more, they become ideal testbeds for benchmarking advanced strategies for implementation and error mitigation, as they have to obey specific, hard constraints: The defining property of a gauge theory is the invariance under local transformations. This means that the generators of the gauge group need to be conserved at each moment in time and space. A simple example with an Abelian gauge group is quantum electrodynamics (QED), where Gauss’s law corresponds to a local $\mathrm{U}(1)$ symmetry that constrains the dynamics of electrically charged matter and the electromagnetic gauge field. More complicated are non-Abelian theories such as quantum chromodynamics (QCD), where a local $\mathrm{SU}(3)$ symmetry governs the interplay of quarks and gluons that can appear in three distinct colors. In contrast to fundamental particles, however, where such conservation laws are postulated as a law of nature, a quantum-simulator device needs to be programmed in such a way as to fulfil it.

In the past, several strategies have been proposed to enforce this local conservation law. One approach is to formulate the Hamiltonian directly using the gauge constraint itself in order to reduce the treated state space to only those configurations that fulfil the conservation law \cite{Martinez2016,Muschik2017,Bernien2017,Klco2018,Kokail2019,Zohar2018,Surace2020}. While this reduces redundancy in the degrees of freedom, integrating out the gauge field can lead to complicated long-range interactions, which can, however, nevertheless be realized in programmable quantum computers \cite{Muschik2017,Martinez2016,Kokail2019}. When integrating out the matter field, theories can be formulated that are purely local \cite{Zohar2018,Surace2020}. Though the resulting interactions can still be quite complicated, experiments on QED in one spatial dimension have been realized exploiting the local blockade between Rydberg atoms \cite{Bernien2017}. 

Despite this progress, it may still be desirable to keep both matter and gauge fields as explicit degrees of freedom in the quantum simulator. First, as Gauss’s law is not exploited at the very inception of the implementation, that approach may provide a larger flexibility in extending the designed building blocks. Second, it enables to probe how gauge invariance can emerge even if a natural system could in principle explore a much larger Hilbert space. This question is also of fundamental relevance to the field of topological matter, which is intrinsically related to gauge theories \cite{Hastings2005,Sachdev2018}. In such a situation, the quantum simulator will contain different Hilbert space sectors that describe different sectors of the gauge symmetry. In principle, any realistic device will have processes that couple these different sectors. Nevertheless, for equilibrium physics, it is well known that gauge theories can emerge at low temperatures even in the presence of Hamiltonian terms that break gauge invariance \cite{Foerster1980,Poppitz2008,Wetterich2017,Bass2020}, which can be understood as a transition into a phase with a Higgs particle of heavy mass \cite{Poppitz2008,Kuno2015,Bazavov2015,HeitgerPhDThesis}. However, for equilibrium physics one can often use powerful quantum Monte Carlo methods \cite{Weinberg_book,Gattringer_book,Cheng_book,Zee_book}, making nonequilibrium situations an even more pertinent target for quantum simulators. 

One promising approach to enforce gauge symmetry even when working far from equilibrium is to exploit global symmetries, such as angular momentum conservation, and promote them to a local symmetry by separating the system into distinct lattice sites \cite{Zohar2013PRA,Stannigel2014,Zache2018}. Such an approach has been realized in a building block using a combination with energetic penalties \cite{Mil2020}. Another promising strategy, which defined much of the initial efforts in the field, is the use of pure energetic penalties to suppress undesired transitions between gauge sectors by rendering them off-resonant. These can be either static terms that are added to the Hamiltonian \cite{Zohar2011,Zohar2012,Zohar2013,Banerjee2012,Hauke2013,Kuehn2014,Kuno2015,Dutta2017,Kuno2017,Negretti2017,Barros2019,Halimeh2020a,Halimeh2020e} or come in the form of dynamical decoupling \cite{Kasper2020}, and they can even correspond to classical dephasing noise \cite{Stannigel2014,Lamm2020}. In all of these, operations are applied that are proportional to the generators of the gauge symmetry, so that potential detrimental transitions between gauge sectors are “rotated away”. An implementation using noisy quantum gates in a digital quantum computer has demonstrated a proof of principle \cite{Samach2021}, and using static energy penalties even a $\mathrm{U}(1)$ lattice gauge theory  of dozens of lattice sites has been realized \cite{Yang2020}. 

Hence, on the one hand, there have been impressive breakthrough results using different implementation strategies, which have even observed relevant many-body phenomena such as Coleman's phase transition \cite{Coleman1976,Yang2020} and have revealed intriguing connections to so-called scar states \cite{Bernien2017,Turner2018}. On the other hand, the vast majority of experimental results was achieved for Abelian lattice gauge theories \cite{Martinez2016,Bernien2017,Dai2017,Klco2018,Kokail2019,Schweizer2019,Goerg2019,Mil2020,Brower2020,Yang2020}. Only a few proofs of principle for non-Abelian theories have been achieved by exploiting the gauge constraint to limit the system to the gauge Hilbert space \cite{Klco2020,Atas2021,Rahman2021,Davoudi2021}. Other implementation strategies still remain unexplored as what concerns experimental realizations of the highly relevant non-Abelian gauge theories, and it still remains a challenge to achieve scaling in system size similar to what has been demonstrated for Abelian gauge theories, while keeping the reliability of the quantum simulator. 

In this work, we study in detail the possibility to retain a non-Abelian gauge symmetry far from equilibrium using only energetic constraints. Our focus is less on a concrete experimental proposal but rather on a more theoretical understanding how well the non-Abelian gauge symmetry can be retained in principle as the dynamics progresses and how that is reflected in the dynamical evolution of physical observables. For concreteness, we focus on a $\mathrm{U}(2)$ gauge theory in the quantum link model formalism \cite{Banerjee2013,Stannigel2014}, although our results are valid for other non-Abelian gauge symmetries, as outlined in our analytic arguments. Using exact diagonalization (ED), we show that gauge protection reliably suppresses gauge violations due to generic gauge-breaking error terms up to all numerically accessible evolution times. These findings are further underpinned using analytic arguments based on methods from periodically driven systems \cite{Abanin2017} and in particular the technique of constrained quantum dynamics \cite{gong2020error,gong2020universal}, which shows an emergent gauge theory reliably reproduces the dynamics up to timescales polynomial in the protection strength. Moreover, we also show that for certain errors a simplified single-body protection term already provides stable gauge invariance up to all accessible evolution times, which we corroborate by analytic arguments based on the quantum Zeno effect \cite{facchi2002quantum,facchi2004unification,facchi2009quantum,burgarth2019generalized,Halimeh2020e,vandamme2021reliability}.
Our work complements a recent study on enforcing non-Abelian gauge symmetry using dynamical decoupling \cite{Kasper2020}.

This article is organized as follows: In Sec.~\ref{sec:model}, we describe the non-Abelian gauge theory and associated gauge-breaking errors. Quench dynamics under full protection are presented in Sec.~\ref{sec:full}. An experimentally feasible single-body protection scheme is then introduced and analyzed in Sec.~\ref{sec:lin}. We conclude and provide future outlook in Sec.~\ref{sec:conc}. We supplement our main text by a description of our ED procedure in~\ref{app:ED}, supporting numerical results in~\ref{app:DiffInitState} for different initial states and values of the model parameters, and a time-dependent perturbation theory derivation in~\ref{app:TDPT}.

\section{$\mathrm{U}(2)$ quantum link model}\label{sec:model}
\begin{figure}[htp]
	\centering
	\includegraphics[width=\textwidth]{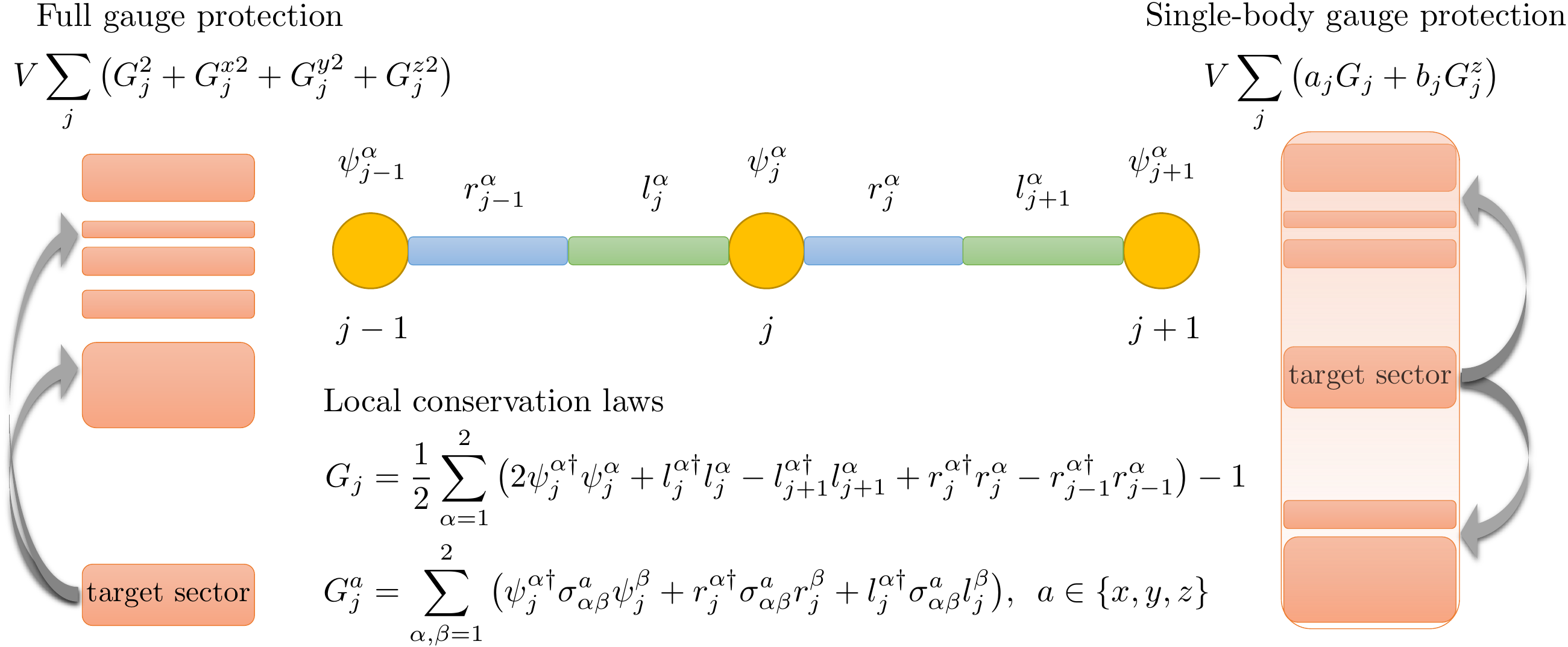}
	\caption{(Color online). Schematic of the non-Abelian lattice gauge theory depicting three matter sites with two links in between. Each matter site hosts two \textit{colors} of matter fields represented by the fermionic annihilation operators $\psi^{1,2}_j$. The gauge link $(j,j+1)$ hosts two colors for each of a left and right rishon, represented by the fermionic annihilation operators $r^{1,2}_j$ and $l^{1,2}_{j+1}$, respectively. The $\mathrm{U}(2)$ quantum link model considered in this work has four different local generators of its non-Abelian gauge symmetry for each matter site $j$ and its two neighboring links: $G_j$, which is the generator of a $\mathrm{U}(1)$ gauge symmetry, and $G_j^a,\,a\in\{x,y,z\}$, which are the three generators of the $\mathrm{SU}(2)$ gauge symmetry. We employ two types of protection in this work: (i) \textit{full} gauge protection, illustrated on the left, which energetically penalizes processes away from the target sector; and (ii) \textit{single-body} gauge protection, illustrated on the right, which renders states in some (but not all) sectors other than the target gauge sector off-resonant, by using single-body terms in $G_j$ and $G^z_j$ with random coefficients $a_j$ and $b_j$. The full protection controllably suppresses violations due to any unitary gauge-breaking errors at sufficiently large volume-independent protection strength $V$ up to all times accessible in exact diagonalization, and is predicted analytically to work up to times exponential in $V$. The single-body protection stabilizes non-Abelian gauge invariance for certain experimentally relevant errors up to all times accessible in exact diagonalization.}
	\label{fig:schematic} 
\end{figure}

We consider a non-Abelian lattice gauge theory (LGT) in the form of the $\mathrm{U}(2)$ quantum link model (QLM) described by the Hamiltonian \cite{Banerjee2013,Wiese_review}
\begin{align}
H_0=&\,\sum_{j=1}^N\left[\sum_{\alpha,\beta=1}^2\Big(J\psi_j^{\alpha\dagger}r_j^\alpha l_{j+1}^{\beta\dagger}\psi_{j+1}^\beta+h\psi_j^{\alpha\dagger}\psi_j^\alpha \psi_{j+1}^{\beta\dagger}\psi_{j+1}^\beta+\text{H.c.}\Big)+\mu\sum_{\alpha=1}^2(-1)^j\psi_j^{\alpha\dagger}\psi_j^\alpha\right],
\end{align}
where $\alpha$ and $\beta$ represent the $\mathrm{U}(2)$ \textit{colors}; see Fig.~\ref{fig:schematic}. The matter fields on site $j$ are represented by the fermionic operators $\psi_j^\alpha$ with a rest mass $m$, while the non-Abelian gauge fields on the link $(j,j+1)$ are denoted by fermionic right and left \textit{rishons} $r_j^\alpha$ and $l_{j+1}^\alpha$, respectively. These fermionic fields satisfy the canonical anticommutation relations
\begin{subequations}
\begin{align}
	\big\{f_m^\alpha,g_n^\beta\big\}&=0,\\
	\big\{f_m^{\alpha\dagger},g_n^\beta\big\}&=\delta_{f,g}\delta_{m,n}\delta_{\alpha,\beta},
\end{align}
\end{subequations}
where $f_j^\alpha,g_j^\alpha\in\big\{\psi^\alpha_j,r^\alpha_j,l^\alpha_j\big\}$.

The non-Abelian gauge invariance of this model is embodied by the relations $[H_0,G_j]=[H_0,G^a_j]=0$ with the four noncommuting generators
\begin{subequations}
	\begin{align}
		G_j&=\frac{1}{2}\sum_{\alpha=1}^2\big(2\psi_j^{\alpha\dagger}\psi_j^\alpha+l_j^{\alpha\dagger}l_j^\alpha-l_{j+1}^{\alpha\dagger}l_{j+1}^\alpha+r_j^{\alpha\dagger}r_j^\alpha-r_{j-1}^{\alpha\dagger}r_{j-1}^\alpha\big)-1,\\
		G^a_j&=\sum_{\alpha,\beta=1}^2\big(\psi_j^{\alpha\dagger}\sigma^a_{\alpha\beta}\psi_j^\beta+r_j^{\alpha\dagger}\sigma^a_{\alpha\beta}r_j^\beta+l_j^{\alpha\dagger}\sigma^a_{\alpha\beta}l_j^\beta\big),
	\end{align}
\end{subequations}
where $a=x,y,z$ and $\sigma^a$ are the Pauli matrices. The $\mathrm{U}(2)$ group symmetry thus separates into a $\mathrm{U}(1)$ part encoded in $G_j$, and an $\mathrm{SU}(2)$ part in $G^a_j$. These generators satisfy the relations $[G_j,G^a_j]=0$ and $[G^a_m,G^b_n] =2i\delta_{m,n}\epsilon_{abc}G^c_n$, with the Levi-Civita tensor $\epsilon_{abc}$. We define the target sector as the set of states $\ket{\phi}$ that satisfy $G_j\ket{\phi}=0,\,G_j^a\ket{\phi}=0,\,\forall j$.

\begin{table}\label{table}\caption{\label{label} Gauge-breaking terms of the error Hamiltonian $H_1$ and their strength in units of $J$ as relevant for the proposal \cite{Stannigel2014}. When two strengths are listed for a given term,the first (second) value holds for even (odd) matter sites. Explicitly, the densities are $n_j^\alpha=\psi^{\alpha\dagger}_j\psi^\alpha_j$, $n_{r,j}^\alpha=r^{\alpha\dagger}_jr^\alpha_j$, and $n_{l,j}^\alpha=l^{\alpha\dagger}_jl^\alpha_j$.}\begin{indented} \item[]\begin{tabular}{@{}llll}
			\br
			process & strength $\lambda$ $[J]$\\
			\mr $\sum_{c=1,2}\psi^{c\dagger}_{j-1}r^c_{j-1}l^{c\dagger}_j\psi^c_j+\text{H.c.}$ & $2.03$\\
			\mr $\sum_{c=1,2}\psi^{c\dagger}_{j-1}r^c_{j-1}l^{\bar{c}\dagger}_j\psi^{\bar{c}}_j+\text{H.c.}$ & $-0.03$\\ 
			\mr $\sum_{c=1,2}\big(\psi^{2\dagger}_jl^1_j\psi^{c\dagger}_{j-1}r^c_{j-1}+l^{c\dagger}_j\psi^c_jr^{1\dagger}_{j-1}\psi^2_{j-1}\big)+\text{H.c.}$ & $0.05$\\
			\mr $\psi^{2\dagger}_jl^1_jr^{1\dagger}_{j-1}\psi^2_{j-1}+\text{H.c.}$ & $1.97$\\
			\mr $\psi^{2\dagger}_j\big(r^2_jr^{1\dagger}_j+l^2_jl^{1\dagger}_j\big)\psi^1_j+\text{H.c.}$ & $0.64,0.6$\\
			\mr $n^1_jn^2_j$ & $0.92,0.74$\\
			\mr $n^1_{r,j}n^2_{r,j}+n^1_{l,j}n^2_{l,j}$ & $1.1,0.88$\\
			\mr $n^1_{r,j}n^2_{l,j+1}+n^2_{r,j}n^1_{l,j+1}$ & $4.21$\\
			\mr $n^1_{r,j}n^1_{l,j+1}$ & $64.77$\\
			\mr $n^2_{r,j}n^2_{l,j+1}$ & $66.54$\\
			\mr $\sum_{c=1,2}n^c_j\big(n^c_{r,j}+n^c_{l,j}\big)$ & $0.18,0.14$\\
			\mr $\sum_{c=1,2}n^c_j\big(n^{\bar{c}}_{r,j}+n^{\bar{c}}_{l,j}\big)$ & $0.84,0.74$\\
			\mr $\sum_{c=1,2}n^c_j\big(n^c_{r,j-1}+n^c_{l,j+1}\big)$ & $0.92$\\
			\mr $n^1_j\big(n^2_{r,j-1}+n^2_{l,j+1}\big)$ & $0.87$\\
			\mr $n^2_j\big(n^1_{r,j-1}+n^1_{l,j+1}\big)$ & $1.16$\\
			\mr $r^{2\dagger}_jr^1_jl^{1\dagger}_{j+1}l^2_{j+1}+\text{H.c.}$ & $59.67$\\
			\mr $\psi^{2\dagger}_j\psi^1_j\big(r^{1\dagger}_{j-1}r^2_{j-1}+l^{1\dagger}_{j+1}l^2_{j+1}\big)+\text{H.c.}$ & $0.39$\\
			\mr $\psi^{1\dagger}_{j+1}l^2_{j+1}r^{2\dagger}_j\psi^1_j+\text{H.c.}$ & $1.91$\\
			\br
	\end{tabular} \end{indented}
\end{table}

\section{Quench dynamics with full gauge protection}\label{sec:full}
Here, we are interested in a potential experimental realization of the ideal theory $H_0$ in, e.g., an ultracold-atom setup. Without infinite fine-tuning, such a quantum-simulation experiment will necessarily lead to gauge-invariance-breaking errors $H_1$. We focus here on the possible errors that for a worst-case scenario have been extensively quantified for the proposal of a $\mathrm{U}(2)$ QLM presented in Ref.~\cite{Stannigel2014}. For completeness, we list these error terms and their corresponding strengths in Table~\ref{table}.

We protect against these errors using the \textit{full} protection term
\begin{align}\label{eq:fullpro}
VH_G=V\sum_{j=1}^N\big(G_j^2+G_j^{x2}+G_j^{y2}+G_j^{z2}\big),
\end{align}
where $V$ is the protection strength. The \textit{faulty} theory,
\begin{align}\label{eq:faulty}
	H=H_0+H_1+VH_G,
\end{align}
is then used to model quench dynamics in an experimentally relevant setup. In Abelian gauge theories, a similar protection term has been shown to penalize processes driving the system away from the target sector, such that a two-regime picture arises: at small $V$, the gauge violation accumulates in an uncontrolled manner, while at sufficiently large $V$, the gauge violation falls in a controlled-error regime where its infinite-time value is $\propto\lambda^2/V^2$, with $\lambda$ the error-strength scale \cite{Halimeh2020a}.

Due to the numerical overhead involved in this problem, and since we are interested in the behavior also at long evolution times, we are restricted in our ED calculations to two matter sites with periodic boundary conditions. This is because each matter site along with its corresponding link are equivalent to six spin-$1/2$ sites (see~\ref{app:ED} for details on the ED implementation). Nevertheless, in view of ongoing ultracold-atom implementations of single building blocks of lattice gauge theories, even in the Abelian case \cite{Schweizer2019,Mil2020}, it is still relevant to benchmark small system sizes. Moreover, as we will discuss later in Sec.~\ref{sec:locobs}, our results for the full protection~\eqref{eq:fullpro} are expected to hold in the thermodynamic limit up to a timescale exponential in the volume-independent protection strength $V$. In this spirit, we consider the two-site initial state
\begin{align}\label{eq:InitState}
	\ket{\phi_0}&=\frac{1}{2}\big(l_1^{1\dagger}\psi_1^{2\dagger}-l_1^{2\dagger}\psi_1^{1\dagger}\big)\big(l_2^{1\dagger}\psi_2^{2\dagger}-l_2^{2\dagger}\psi_2^{1\dagger}\big)\ket{0},
\end{align}
which, being a product state, is relatively simple to implement in an experiment. Moreover, it is in the target sector since it satisfies $G_j\ket{\phi_0}=G_j^a\ket{\phi_0}=0,\,\forall j$. As shown in \ref{app:DiffInitState}, other initial states will not alter the qualitative behavior of what is discussed in the following.

\subsection{Gauge violation}\label{sec:viol}
We start by numerically calculating the quench dynamics of the temporally averaged gauge violation 
	\begin{align}\label{eq:viol}
	\varepsilon=&\frac{1}{Nt}\int_0^tds\,\bra{\phi(s)}H_G\ket{\phi(s)},  
	\end{align}
where $\ket{\phi(t)}=e^{-iHt}\ket{\phi_0}$. 	
In Sec.~\ref{sec:locobs}, we will study other local observables, for which we provide also further analytic arguments valid in the thermodynamic limit.

\begin{figure}[htp]
	\centering
	\includegraphics[width=.75\textwidth]{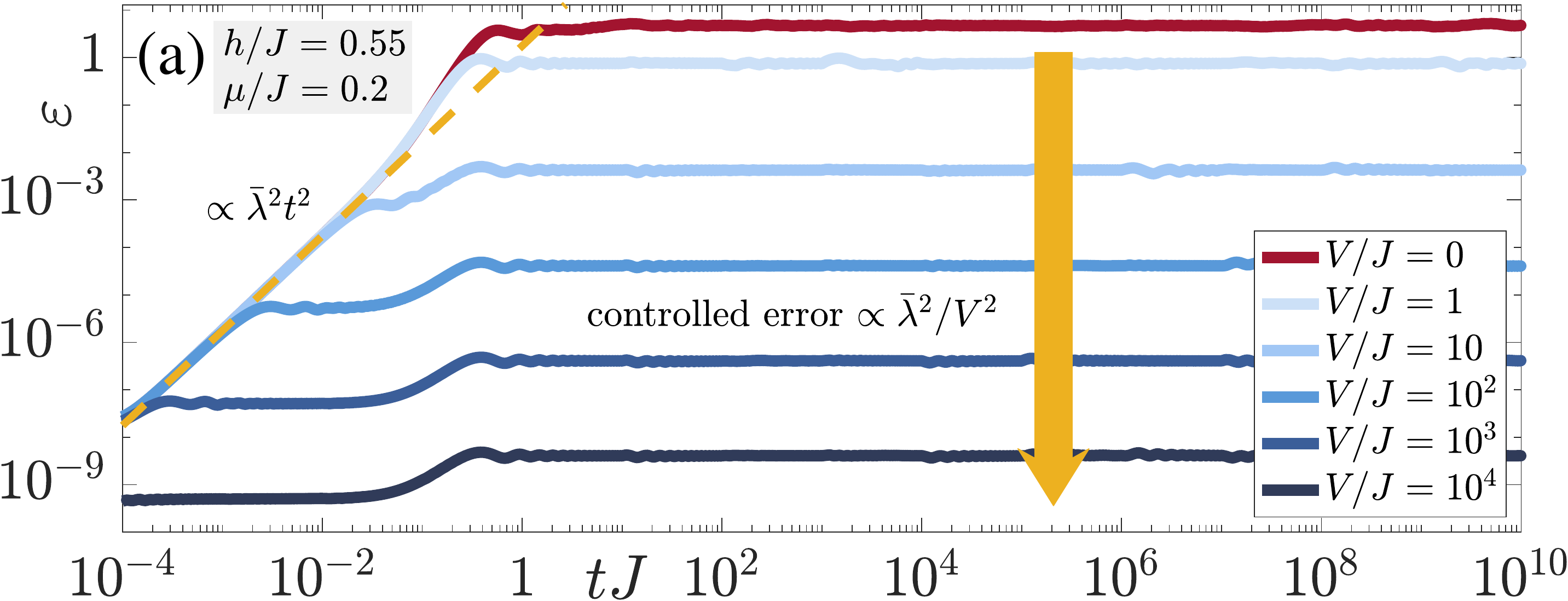}\\
	\vspace{1.5mm}
	\includegraphics[width=.75\textwidth]{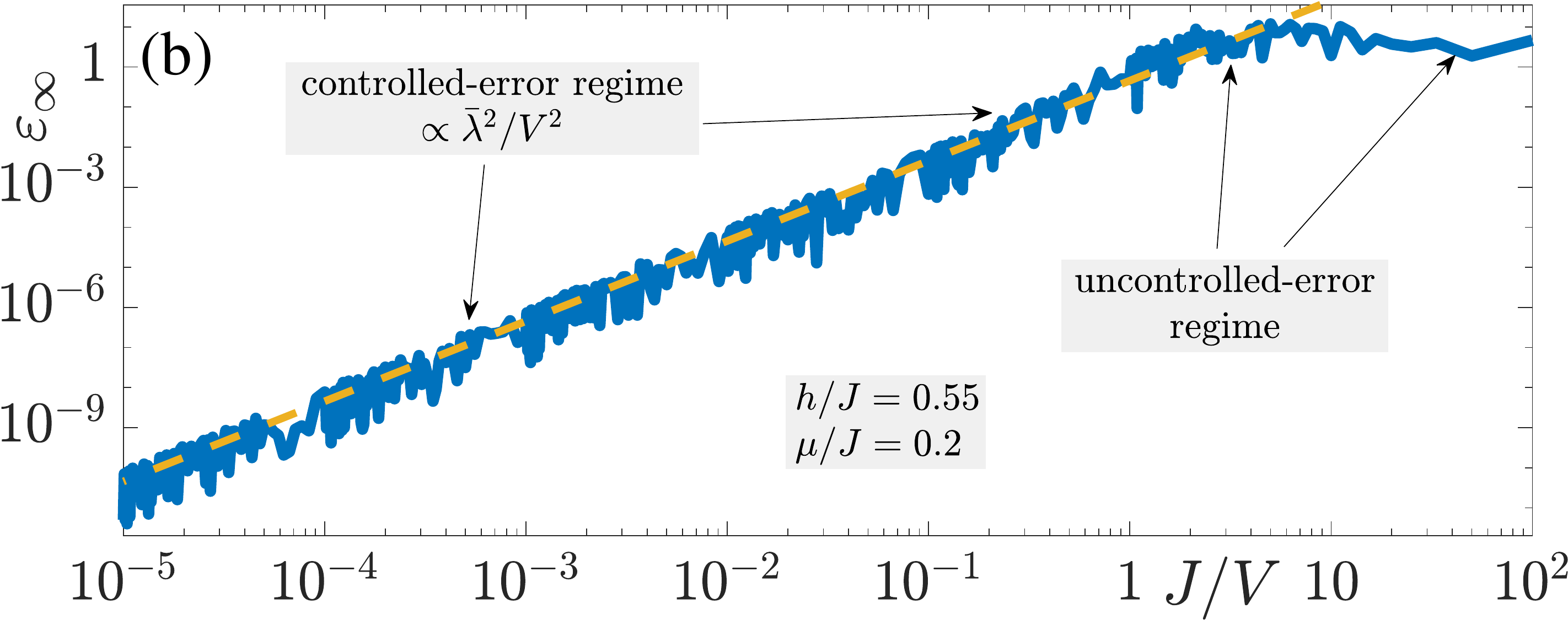}
	\caption{(Color online). Dynamics of the gauge violation in the wake of quenching a gauge-invariant initial state $\ket{\phi_0}$ in the physical sector by the Hamiltonian of a faulty theory, $H=H_0+H_1+VH_G$. (a) The temporally averaged gauge violation shows reliable suppression even at very small values of the protection strength $V\sim J$, where at long times the gauge violation enters a plateau $\propto\bar{\lambda}^2/V^2$. (b) The ``infinite-time'' gauge violation as a function of $J/V$. Two clear regimes appear, one of controlled error for $V\gtrsim J$ and a second of uncontrolled error $\propto\bar{\lambda}^2/V^2$ for $V\lesssim J$. We have checked that our results are valid for different initial states and generic values of the microscopic parameters.}
	\label{fig:violation} 
\end{figure}

We quench the initial state of Eq.~\eqref{eq:InitState} by the faulty-theory Hamiltonian $H$ of Eq.~\eqref{eq:faulty} with the gauge protection $VH_G$ given in Eq.~\eqref{eq:fullpro}. The ensuing dynamics of the gauge violation is shown in Fig.~\ref{fig:violation}(a) for a gauge-breaking term $H_1$ including all the processes in Table~\ref{table}, and at various values of the protection strength $V$. Importantly, the error $H_1$ is not perturbative, with some of its terms carrying strengths $\lambda\gg J>0$. The gauge violation initially grows $\propto\bar{\lambda}^2t^2$ at short times, which can be derived in time-dependent perturbation theory (TDPT); cf.~\ref{app:TDPT}. Here, we have quantified the overall strength of $H_1$ by the average $\bar{\lambda}$ of all $\lambda$ values in Table~\ref{table}. Without any protection ($V=0$), this behavior persists until a timescale $\propto1/\bar{\lambda}$. Beyond this timescale, the gauge violation settles into a steady state of  \textit{maximal violation}. However, once the protection strength $V$ is sufficiently large, a plateauing behavior at a timescale $\propto1/V$ supersedes that $\propto1/\bar{\lambda}$. This new timescale is when the protection term becomes dominant in the dynamics. The resulting gauge-violation plateau persists for all accessible evolution times in ED, and it lies at a lower value $\propto\bar{\lambda}^2/V^2$. This behavior is qualitatively similar to its counterpart in the case of Abelian gauge theories, and can for finite systems also be analytically derived in degenerate perturbation theory \cite{Halimeh2020a} (see also the analytical discussion in Sec.~\ref{sec:locobs}). Importantly, already at rather small values of $V\approx10J$ we find reliable and controlled suppression of the gauge violation despite the nonperturbative error $H_1$.

It is informative to study the ``infinite-time'' gauge violation as a function of $J/V$, in order to see if a transition will arise between a regime of \textit{controlled} and \textit{uncontrolled error} as has been found for Abelian gauge theories under various protection schemes \cite{Halimeh2020a,Halimeh2020e}. Indeed, and as shown in Fig.~\ref{fig:violation}(b), once $V$ is sufficiently large, the infinite-time violation shows a quadratic behavior $\propto\bar{\lambda}^2/V^2$. This indicates a controlled-error regime in which the faulty dynamics can be smoothly connected to those under the ideal theory $H_0$. In contrast, when $V$ is sufficiently small, the infinite-time gauge violation displays a chaotic behavior insomuch that its value exhibits no clear relation to $V$. This is very similar to the chaos--quantum-localization transition one finds in the Trotterization error of local observables in the dynamics of a digital quantum-simulator device \cite{Heyl2019}. There, the long-time error in a local observable can be shown to scale $\propto\tau^2$ when the Trotter time-step $\tau$ is below a critical value, whereas above it the long-time error shows chaotic behavior that cannot be retrieved from the value of $\tau$. In Ref.~\cite{Heyl2019}, this quantum localization has been related to the phase transition between a many-body localized phase (controlled-error regime) and a quantum chaotic phase (uncontrolled-error regime). From Fig.~\ref{fig:violation}(b), one can deduce a possible transition point at $V_\text{c}\sim \mathcal{O}(J)$. This is a powerful result showing that one can implement the full gauge protection scheme of Eq.~\eqref{eq:fullpro} at a strength of the order of the coupling constant $J$ and still achieve dynamics that can be analytically connected to the ideal case, at least for finite system sizes. It is worth noting that this putative dynamical critical point depends on the initial state (see discussion in~\ref{app:DiffInitState}).

\begin{figure}[htp]
	\centering
	\includegraphics[width=.75\textwidth]{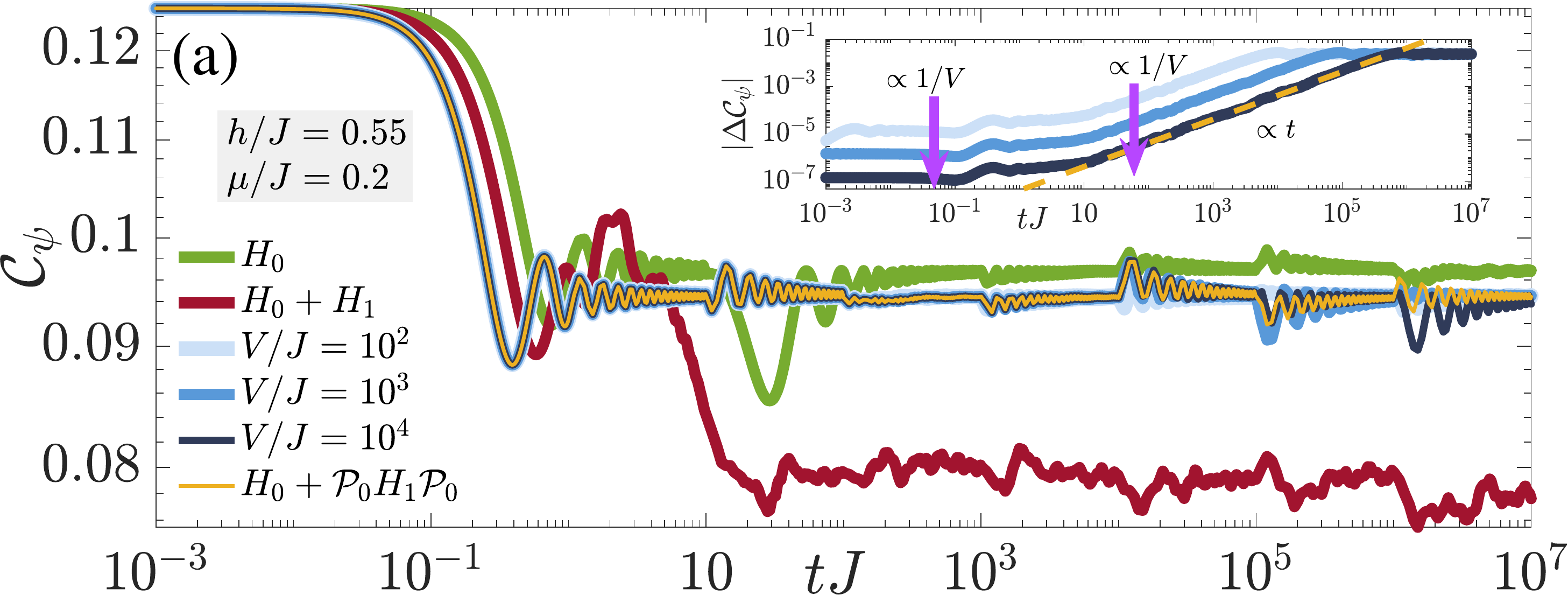}\\
	\vspace{1.5mm}
	\includegraphics[width=.75\textwidth]{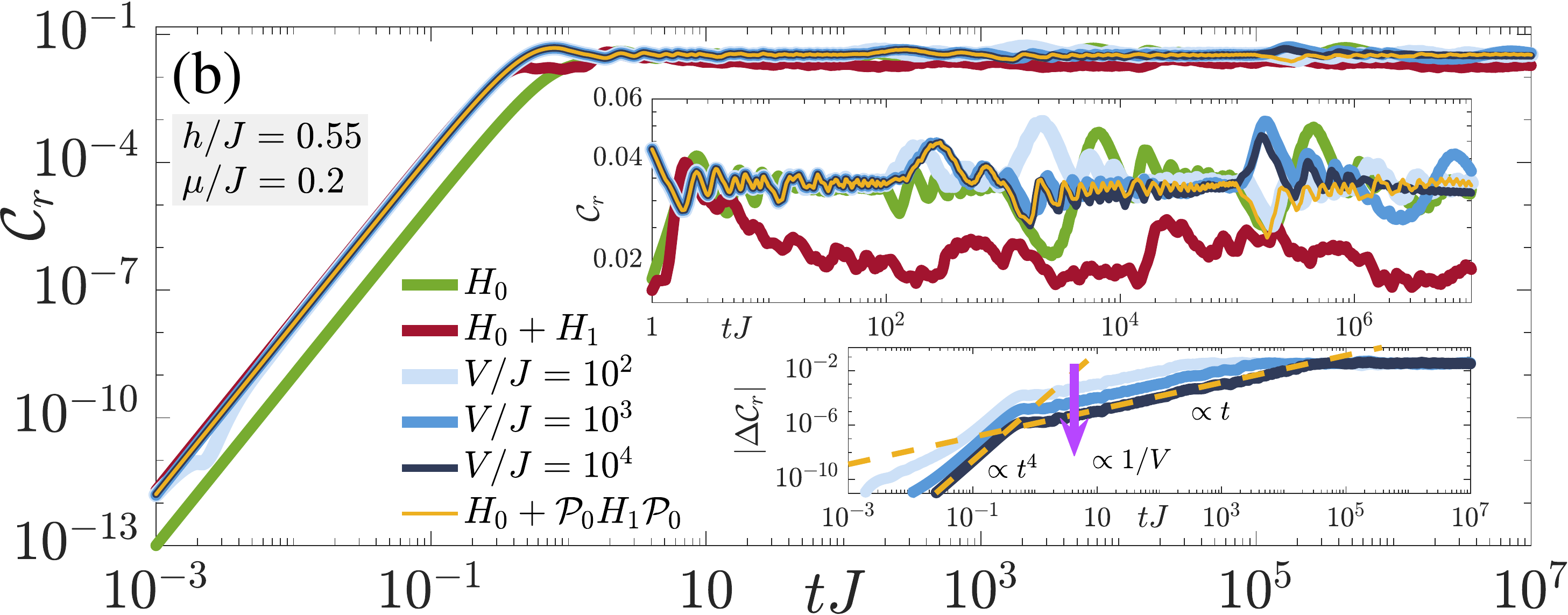}\\
	\vspace{1.5mm}
	\includegraphics[width=.75\textwidth]{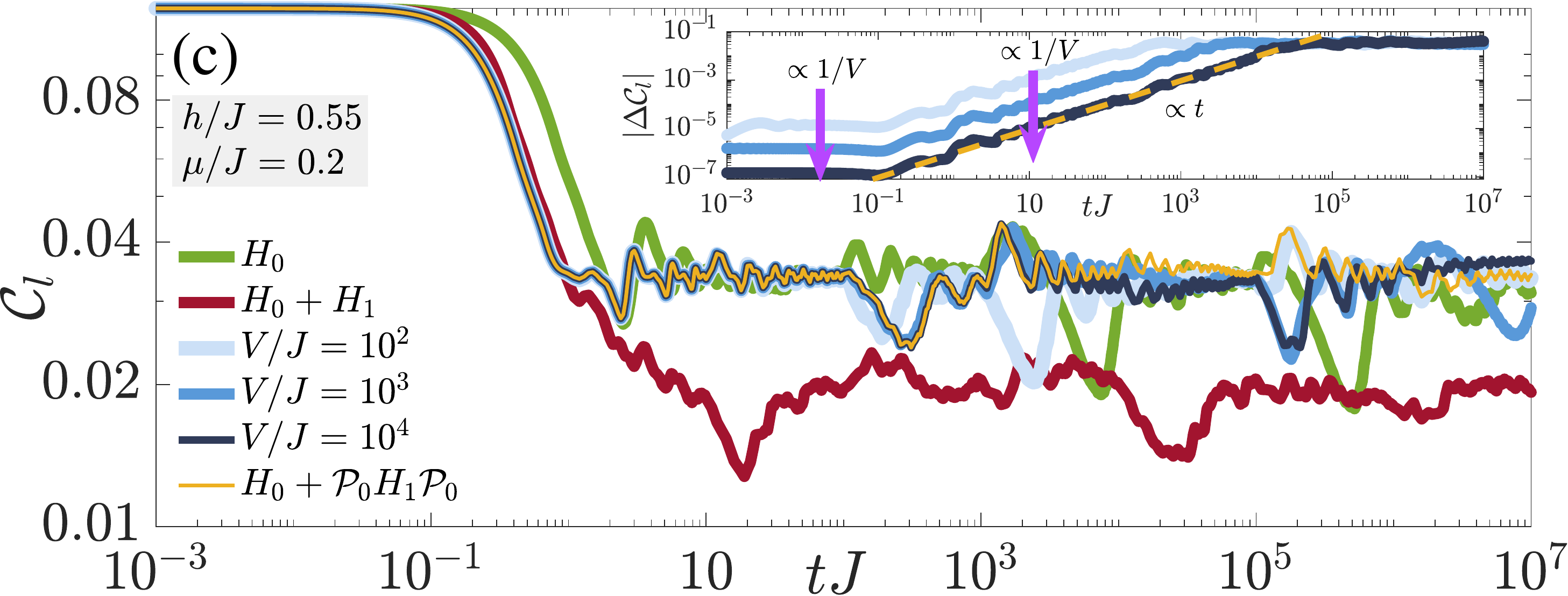}
	\caption{(Color online). Temporally averaged connected density-density correlations of the (a) matter fields, and the (b) right and (c) left rishons. In each panel we show the ideal gauge theory dynamics (green curve), the dynamics under $H_0+H_1$ (red curve), the dynamics under the faulty theory $H=H_0+H_1+VH_G$ at $V/J=10^2,\,10^3,\,10^4$ (different shades of blue), and the dynamics under the adjusted gauge theory $H_\text{adj}=H_0+\mathcal{P}_0H_1\mathcal{P}_0$ (yellow curve). Even though without any protection ($V=0$) the faulty theory gives rise to significantly different dynamics than the ideal theory, upon introducing gauge protection we find that the dynamics is well reproduced by the adjusted gauge theory. Indeed, the insets show that the error is suppressed by $V$, as predicted analytically, and grows milder than the analytic prediction of $\propto t^2$. Even though the adjusted timescale is analytically derived to be $\propto\sqrt{V/V_0^3}$, this is an earliest estimate in a worst-case scenario, and our numerical results indicate that it is $\propto V/J^2$.}
	\label{fig:observables} 
\end{figure}

\subsection{Local observables and analytic arguments}\label{sec:locobs}

A highly relevant question is in how far the controlled dynamics of the gauge violation is reflected in other local observables. In Refs.~\cite{Halimeh2020e,vandamme2021reliability} it has been shown in the case of full protection that the dynamics of local observables under the faulty theory can be reproduced by two different emergent gauge theories up to corresponding error bounds. Even though this result has been derived for Abelian gauge theories, it remains the same for their non-Abelian counterparts, as detailed below. All the emergent gauge theories considered in this work are based on the assumption of initial states prepared in the target sector, although these theories can readily be extended to initial states starting in any gauge-invariant sector. 

The first emergent theory, whose derivation relies mainly on the technique of ``contrained quantum dynamics'' developed by Gong \textit{et al.} \cite{gong2020error,gong2020universal}, is the \textit{adjusted} gauge theory given by $H_\text{adj}=H_0+\mathcal{P}_0H_1\mathcal{P}_0$, where $\mathcal{P}_0$ is the projector onto the target subspace $\{\ket{\phi}\}$ such that $G_j\ket{\phi}=G^x_j\ket{\phi}=G^y_j\ket{\phi}=G^z_j\ket{\phi}=0,\,\forall j$. The method of contrained quantum dynamics requires that (i) there is a sufficiently large gap between the ground state and excited states of the protection Hamiltonian, (ii) the protection Hamiltonian $VH_G$ can be expressed as the summation of local commuting terms, which are $G_j^2 + G_j^{x2} + G_j^{y2} + G_j^{z2} $ in our case, and (iii) the ground state of $VH_G$ also minimizes the energy of local commuting terms (frustration-free condition). Then, the dynamics of a local observable $O$ under the faulty theory is reproduced by the adjusted gauge theory within the error bound \cite{gong2020error,gong2020universal,vandamme2021reliability}
\begin{align}
	\label{eq:AdjError}
	\Big\lvert\bra{\phi_0} e^{iHt}Oe^{-iHt} - e^{iH_\text{adj}t}Oe^{-iH_\text{adj}t}\ket{\phi_0}\Big\rvert \leq \Delta_\text{adj},
\end{align}
with the upper error bound $\Delta_\text{adj}\sim t^2V_0^3/V$, and where $V_0$ is an energy scale depending on the microscopic parameters of the faulty theory \cite{vandamme2021reliability}. Crucially, this error bound does not depend on system size, from which robustness of non-Abelian gauge invariance is to be expected also in the thermodynamic limit. From Eq.~\eqref{eq:AdjError}, we can deduce that the adjusted gauge theory $H_\text{adj}$ will be able to controllably reproduce the dynamics of the faulty gauge theory $H$ up to the timescale $\tau_\text{adj}\propto\sqrt{V/V_0^3}$ at the earliest.

In order to study the emergence of the adjusted gauge theory governing local observables, we plot in Fig.~\ref{fig:observables} the dynamics of the density-density correlations 
\begin{align}
	\label{eq:corr}
	\mathcal{C}_f=&\frac{1}{N^2t}\sum_{m,n=1}^N\int_0^tds\,\big[\bra{\phi(s)}f^{1\dagger}_m f^1_m f^{2\dagger}_n f^2_n\ket{\phi(s)}-\bra{\phi(s)}f^{1\dagger}_m f^1_m\ket{\phi(s)}\bra{\phi(s)}f^{2\dagger}_n f^2_n\ket{\phi(s)}\big],
	\end{align}
where $f_j^\alpha\in\big\{\psi^\alpha_j,r^\alpha_j,l^\alpha_j\big\}$. The conclusions remain unaltered regardless of whether the correlation is that of the matter fields or rishons. The ideal-theory dynamics (green curve) and those under $H_0+H_1$ (red) deviate significantly from each other starting at very early times. However, at sufficiently large protection strength $V$, we see that the dynamics (different shades of blue) is faithfully reproduced by the adjusted gauge theory $H_\text{adj}$ up to a timescale $\propto V/J^2$, which is longer than $\tau_\text{adj}\propto\sqrt{V/V_0^3}$ predicted analytically \cite{vandamme2021reliability}. In the present case, the adjusted gauge theory does not coincide with the ideal gauge theory $H_0$, because $\mathcal{P}_0H_1\mathcal{P}_0\neq0$ since $H_1$ contains processes within the target sector $G_j\ket{\psi}=G_j^a\ket{\psi}=0,\,\forall j$. In cases when $H_1$ contains no such processes, the adjusted gauge theory is identical to $H_0$ \cite{vandamme2021reliability}. Importantly, however, even when $H_\text{adj}$ and $H_0$ yield different dynamics, both cases are exact gauge theories. As such, even though experimentally one may not be able to reproduce the desired ideal-theory dynamics, one is still able to implement the dynamics of an adjusted gauge theory with the same local gauge symmetry. Even more, in some cases one may be interested in studying precisely the additional processes generated by $\mathcal{P}_0H_1\mathcal{P}_0$.
 
When $V$ is sufficiently large, there emerges also a renormalized gauge theory $H_\text{ren}$ that governs the dynamics of a local observable $O$ up to a timescale $\propto\exp(V/V_0)/V_0$ within the error bound \cite{vandamme2021reliability,abanin2017rigorous}
 \begin{align}
 	\label{eq:AdjError2}
 	\Big\lvert\bra{\phi_0} e^{iHt}Oe^{-iHt} - e^{iH_\text{ren}t}Oe^{-iH_\text{ren}t}\ket{\phi_0}\Big\rvert \leq K(O)/V_0,
\end{align}
where $ K(O)$ is model parameter-dependent but volume and $V$-independent \cite{vandamme2021reliability}. Unlike the adjusted theory, the renormalized theory requires that (i) the spectrum of each local commuting term is comprised of integers, and that (ii) the kernel of $VH_G$ is exactly the target sector. Non-Abelian gauge theories with full protection satisfy both conditions. The general form of $H_\text{ren}$ is hard to obtain, but it is still a gauge theory with the same local gauge symmetry as the ideal model. The renormalized gauge theory also works in the thermodynamic limit at a volume-independent protection strength $V$, and dominates the dynamics after the adjusted gauge theory breaks down.
 
 \section{Quench dynamics with single-body protection}\label{sec:lin}
 
 \begin{figure}[htp]
 	\centering
 	\includegraphics[width=.75\textwidth]{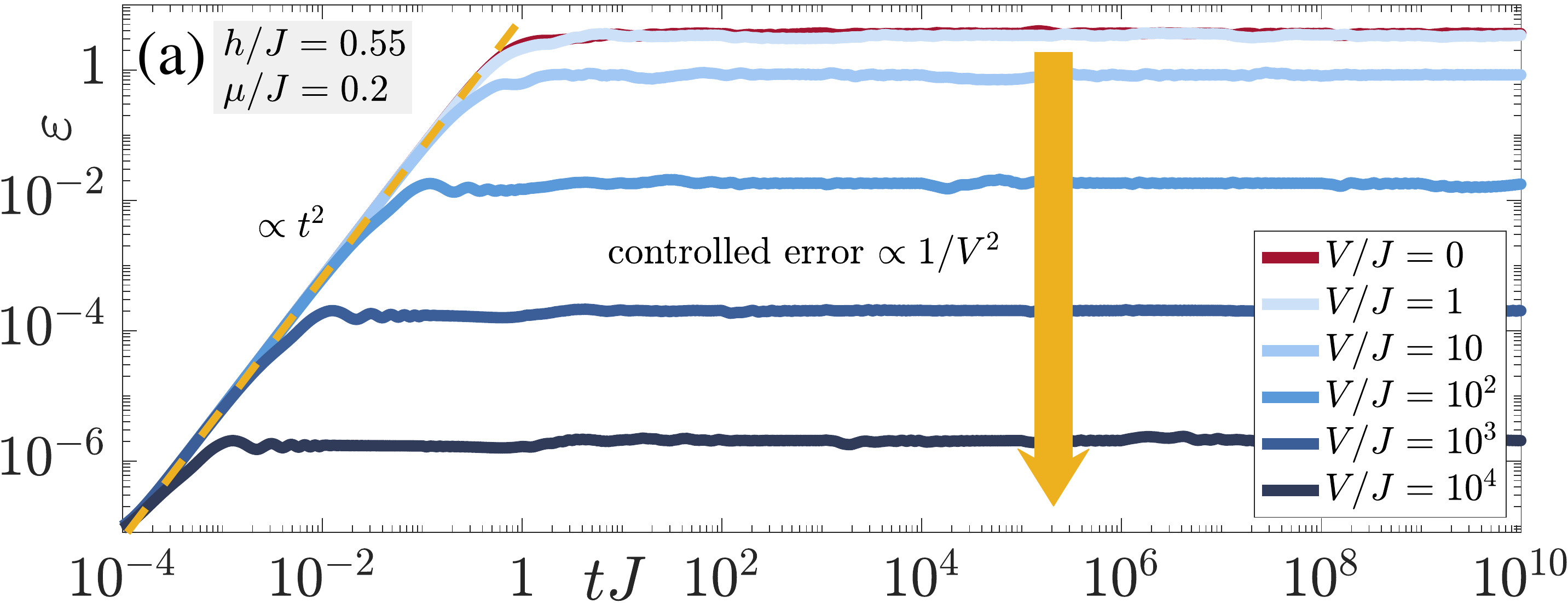}\\
 	\vspace{1.5mm}
 	\includegraphics[width=.75\textwidth]{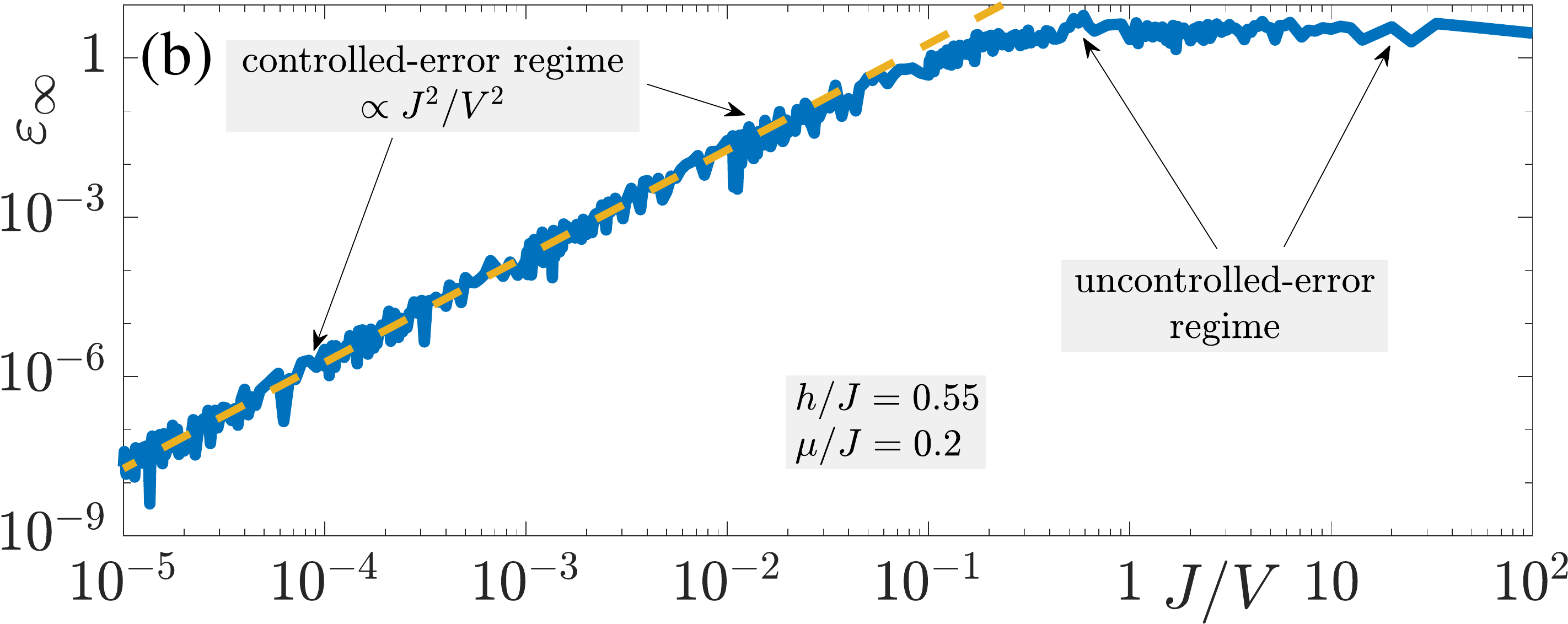}
 	\caption{(Color online). Dynamics of the gauge violation in the wake of quenching a gauge-invariant initial state $\ket{\phi_0}$ in the physical sector by the faulty-theory Hamiltonian $H=H_0+\tilde{H}_1+V\tilde{H}_G$, where the gauge protection is now composed of only single-body terms; see Eq.~\eqref{eq:LinPro}. (a) The temporally averaged gauge violation shows reliable suppression at sufficiently large values of $V$, where at long times the gauge violation enters a plateau $\propto1/V^2$. (b) ``Infinite-time'' gauge violation as a function of $J/V$. Two clear regimes appear, one of controlled error $\propto1/V^2$ for $V\gtrsim J$ and a second one of uncontrolled error for $V\lesssim 10J$. We have checked that our results are valid for different initial states and generic values of the microscopic parameters.}
 	\label{fig:violation_LP} 
 \end{figure}

\begin{figure}[htp]
\centering
\includegraphics[width=.75\textwidth]{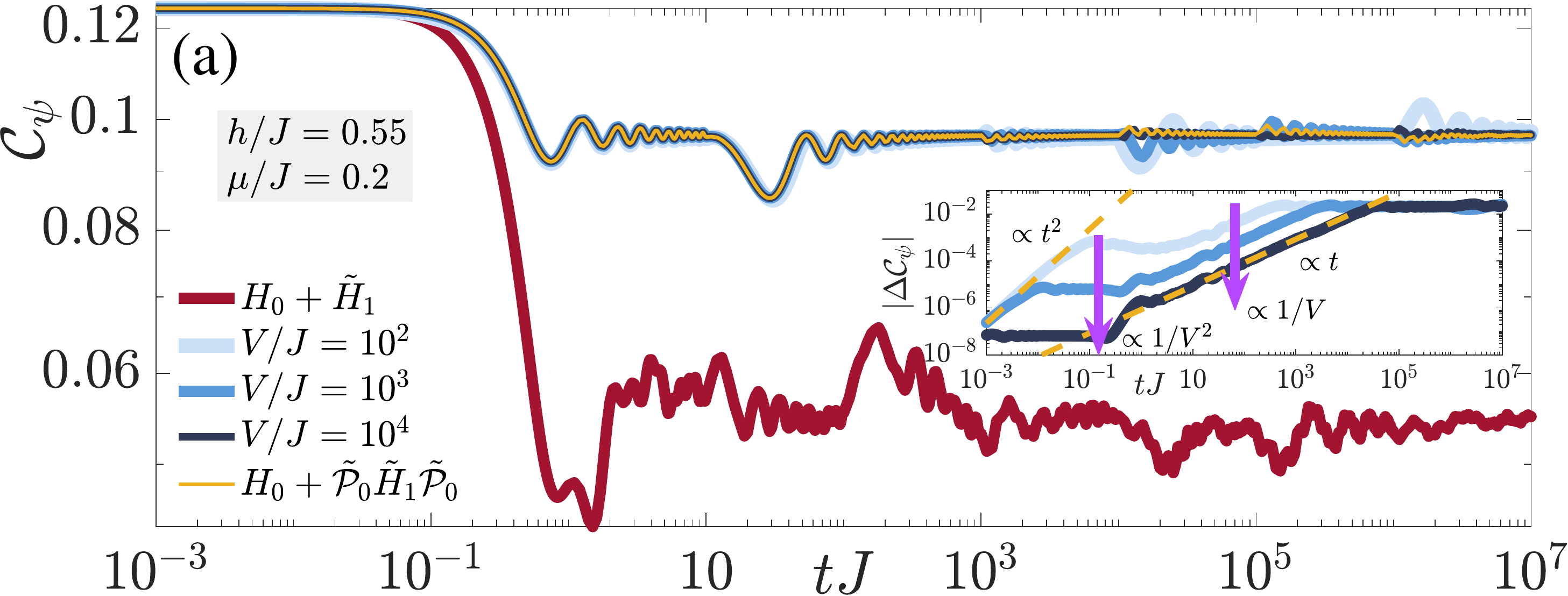}\\
\vspace{1.5mm}
\includegraphics[width=.75\textwidth]{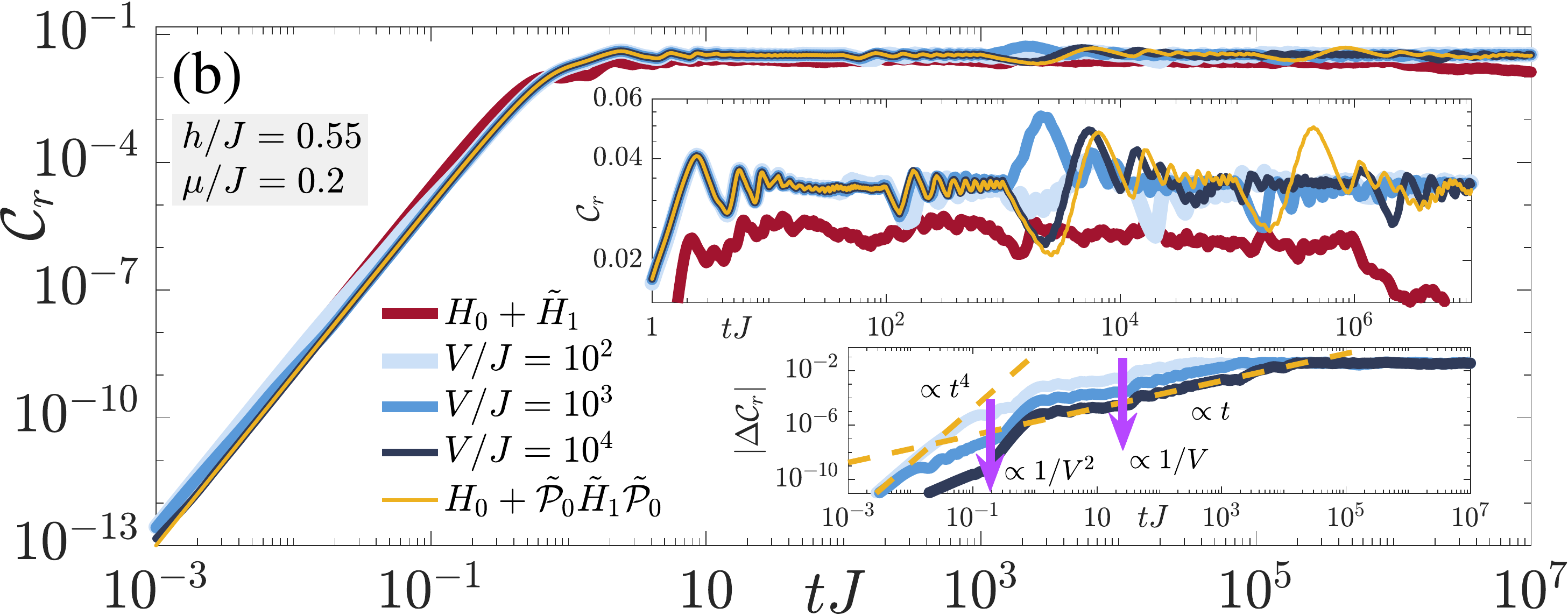}\\
\vspace{1.5mm}
\includegraphics[width=.75\textwidth]{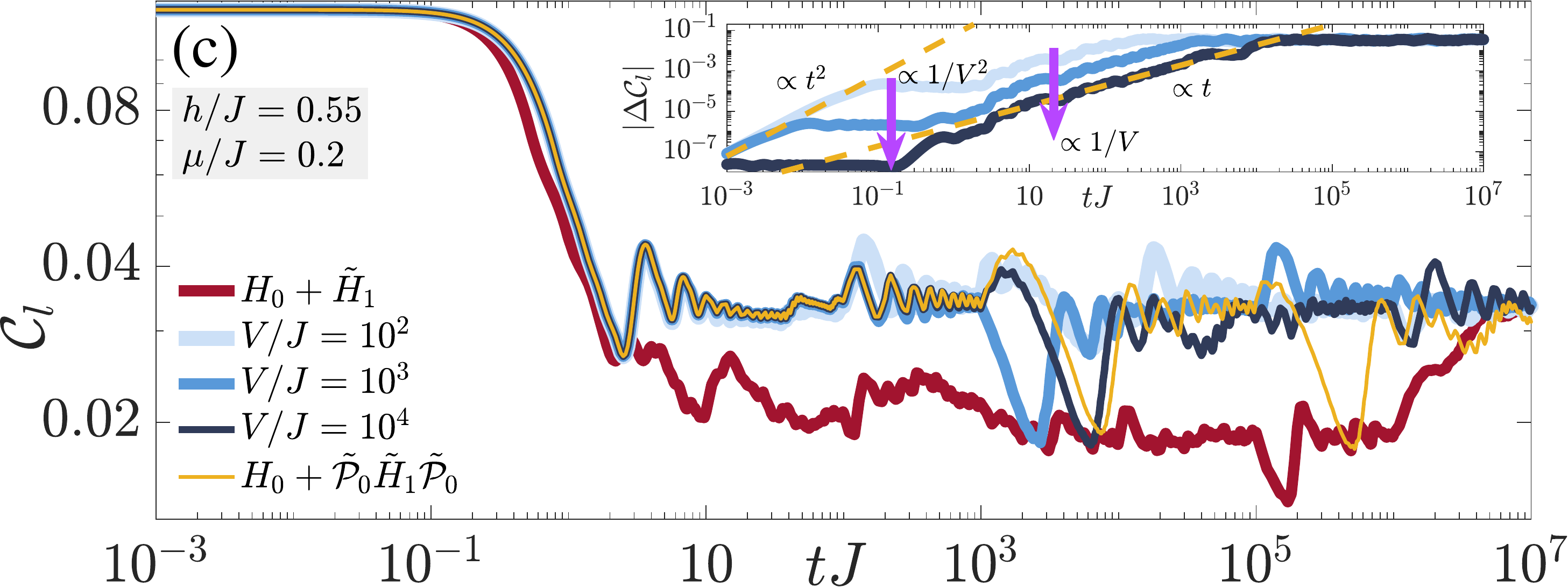}
\caption{(Color online). Temporally averaged connected density-density correlations of (a) the matter fields, and (b) the right and (c) left rishons. In each panel, we show the ideal gauge theory dynamics (yellow curve), which in this case is identical to the adjusted gauge theory since $\tilde{\mathcal{P}}_0\tilde{H}_1\tilde{\mathcal{P}}_0=0$, the dynamics under $H_0+H_1$ (red), and the dynamics under the faulty theory $H=H_0+\tilde{H}_1+V\tilde{H}_G$ at $V/J=10^2,\,10^3,\,10^4$ (different shades of blue). Even though without any protection ($V=0$) the faulty theory gives rise to significantly different dynamics than the ideal theory, upon introducing gauge protection the dynamics resembles that of the ideal theory. As the insets show, the error is suppressed by $V$, grows very slowly in time, and the timescale of the adjusted gauge theory is $\propto V/J^2$.}
\label{fig:observables_LP} 
\end{figure}

 Recently, it has been shown that one can protect gauge invariance in Abelian $\mathrm{U}(1)$ gauge theories through a \textit{single-body protection} term of the form $V\sum_ja_jG_j$, where $G_j$ is the generator of the local $\mathrm{U}(1)$ gauge symmetry and $a_j$ are real numbers normalized such that $\max\{\lvert a_j\rvert\}=1$ and chosen such that $\sum_ja_jg_j=0$ if and only if $g_j=0,\,\forall j$, with $g_j$ the eigenvalues of $G_j$ \cite{Halimeh2020e}. In the case of non-Abelian gauge theories, such a scheme is faced with the problem of noncommuting local generators, as is the case with the $\mathrm{U}(2)$ QLM. This model has the generator $G_j$ for the local $\mathrm{U}(1)$ gauge symmetry and the generators $G_j^a$ for the local $\mathrm{SU}(2)$ gauge symmetry, with $[G_j^a,G_j^b]=2i\epsilon_{abc}G_j^c$. This renders the kernel of the linear summation of these generators no longer the physical target sector, which makes single-body protection generically inadequate for the emergence of a renormalized gauge theory, unlike in the case of full protection. Furthermore, when it comes to the emergence of an adjusted gauge theory as derived through constrained quantum dynamics, the physical sector is not the ground state of the single-body protection Hamiltonian, and, as such, constrained quantum dynamics cannot be applied here.
 
 Nevertheless, for simplified errors, it may be possible to make use of single-body protection to produce the adjusted gauge theory through the quantum Zeno effect (QZE) \cite{facchi2002quantum,facchi2004unification,facchi2009quantum,burgarth2019generalized,Halimeh2020e,vandamme2021reliability}. For this purpose, we consider the simplified error term
 \begin{align}\label{eq:ColorChanging}
 	\tilde{H}_1=\sum_{j=1}^N\bigg[\psi_j^{2\dagger}r_j^1l_{j+1}^{1\dagger}\psi_{j+1}^2+\sum_{\alpha=1,2}\big(r_j^{\alpha\dagger}\psi_j^\alpha l_{j+1}^{1\dagger}\psi_{j+1}^2+\psi_j^{2\dagger}r_j^1\psi_{j+1}^{\alpha\dagger}l_{j+1}^\alpha\big)+\text{H.c.}\bigg],
 \end{align}
which represents gauge-breaking color-changing processes. To suppress gauge violations due to these terms, we employ the single-body protection 
\begin{align}\label{eq:LinPro}
	V\tilde{H}_G=V\sum_{j=1}^N\big(a_jG_j+b_jG_j^z\big),
\end{align}
where $a_j$ and $b_j$ are real random numbers in $[-1,1]$. Note how only one $\mathrm{SU}(2)$ generator, $G_j^z$, is included. The restrictions on the QZE-based adjusted gauge theory are much looser than either its counterpart based on constrained quantum dynamics or the renormalized gauge theory. It only requires that $\tilde{\mathcal{P}}_0\tilde{H}_1\tilde{\mathcal{P}}_0$ is gauge-invariant, where $\tilde{\mathcal{P}}_0$ is the projector onto the eigenstates of $\tilde{H}_G$ with zero eigenvalue, and yields the adjusted gauge theory $H_0 + \tilde{\mathcal{P}}_0\tilde{H}_1\tilde{\mathcal{P}}_0$. In principle, the QZE-based adjusted gauge theory can analytically only guarantee a worst-case scenario of a volume-dependent protection timescale $\tau_\text{adj}\propto V/(V_0N)^2$. Nevertheless, the iMPS results in Ref.~\cite{vandamme2021reliability} show the protection timescale for certain Abelian gauge theories is likely volume-independent, at least for local observables.

In what follows, we quench the initial state in Eq.~\eqref{eq:InitState} with $H=H_0+\tilde{H}_1+V\tilde{H}_G$ and study the dynamics of the gauge violation~\eqref{eq:viol} and the density-density correlations~\eqref{eq:corr}. 
The time evolution of the gauge violation is shown in Fig.~\ref{fig:violation_LP}(a) for various values of $V$ at $h/J=0.55$ and $\mu/J=0.2$. We have checked that our conclusions hold for other generic values of $h$ and $\mu$ and for different initial states (see~\ref{app:DiffInitState}). 
As in the case of full protection discussed in Sec.~\ref{sec:viol}, the gauge violation enters a plateau $\propto1/V^2$ at a timescale $\propto1/V$ when $V$ is large enough, after an initial growth $\propto t^2$ at early times in agreement with TDPT (see~\ref{app:TDPT}). The infinite-time gauge violation is shown in Fig.~\ref{fig:violation_LP}(b) also for $h/J=0.55$ and $\mu/J=0.2$ as a function of $J/V$. Qualitatively, the conclusion is identical to that in the case of full gauge protection. At sufficiently small values of $V$, the infinite-time violation is uncontrolled, i.e., one cannot extract its value from that of $V$. However, it enters a controlled-error regime at sufficiently large $V$, where it scales $\sim J^2/V^2$. This result is impressive considering that the single-body protection terms in Eq.~\eqref{eq:LinPro} can be readily implemented with much smaller overhead than the ideal gauge theory itself. This feature can be very useful for ongoing efforts to realize and stabilize non-Abelian lattice gauge theories in quantum synthetic matter setups.

The dynamics of the density-density correlations in Fig.~\ref{fig:observables_LP} show qualitatively similar behavior to the case of full protection, with the minor exception that in the case of the gauge-breaking error $\tilde{H}_1$ given in Eq.~\eqref{eq:ColorChanging}, the ideal gauge theory $H_0$ is itself the adjusted gauge theory since $\tilde{\mathcal{P}}_0\tilde{H}_1\tilde{\mathcal{P}}_0=0$. Once again, we find that the dynamics under the faulty theory $H$ is adequately reproduced up to a timescale $\propto V/J^2$, but due to our small system size, we cannot ascertain whether this is longer than the analytically predicted timescale $\tau_\text{adj}\propto V/(V_0N)^2$ \cite{Halimeh2020e,vandamme2021reliability}.

\section{Conclusion and outlook}\label{sec:conc}
We have shown that faulty non-Abelian gauge theories with generic nonperturbative gauge-breaking errors can be reliably stabilized up to indefinite times in exact diagonalization using \textit{full} gauge protection based on energy penalties. We have also shown that an adjusted gauge theory arises that reproduces the dynamics of local observables in the faulty theory up to a timescale proportional to a volume-independent protection strength. In addition, we have presented rigorous analytic arguments predicting the adjusted gauge theory, in addition to an emergent \textit{renormalized} gauge theory that reproduces the dynamics of local observables in the faulty theory up to a timescale exponential in a volume-independent protection strength. As such, even though our exact diagonalization calculations are restricted to two matter sites due to numerical overhead, we expect energetic gauge protection to be a viable error-mitigation scheme for non-Abelian gauge theories also in the thermodynamic limit.

Moreover, we have introduced a single-body protection scheme for non-Abelian gauge theories that is simple to implement with standard experimental capacities of manipulating single lattice sites. For a certain class of nonperturbative errors, our exact diagonalization results show that this simple protection term  reliably suppresses gauge violations up to all accessible evolution times. An interesting future avenue may be to generalize the single-body protection scheme to handle other generic unitary errors. Another important question is how well the gauge protection works in more than one spatial dimension \cite{Zohar2021quantum}. The independence of our analytic arguments of dimensionality allows us to anticipate that there is a certain degree of robustness even in higher-dimensional systems.

\section*{Acknowledgments}
The authors are grateful to Debasish Banerjee, Marcello Dalmonte, Federica M.~Surace, and Maarten Van Damme for valuable comments. This work is part of and supported by Provincia Autonoma di Trento, the ERC Starting Grant StrEnQTh (project ID 804305), the Google Research Scholar Award ProGauge, and Q@TN — Quantum Science and Technology in Trento.

\appendix
\section{Jordan-Wigner transformation for multi-species fermionic models}\label{app:ED}
Consider a chain of $N$ sites, each of which can host one of $p$ different species of fermions. We denote the annihilation operator of each fermion as $c_{s,j}$ where $j=1,\ldots,N$ is the site index and $s=1,\ldots,p$ denotes the ``spin'' degree of freedom of the corresponding fermionic species. We first build the associated Pauli spin basis $\sigma_{s,j}$, which comprises a total of $pN$ spins. Accordingly, the Jordan--Wigner transformation that gives us the fermionic operators $c_{s,j}$ is
\begin{subequations}\label{eq:JW}
	\begin{align}
		c_{1,j}&=\bigg(\prod_{l<j}\prod_s\sigma_{s,l}^z\bigg)\sigma_{1,j}^-,\\
		c_{2,j}&=\bigg(\prod_{l<j}\prod_s\sigma_{s,l}^z\bigg)\bigg(\prod_{s'=1}^1\sigma_{s',j}^z\bigg)\sigma_{2,j}^-,\\
		c_{3,j}&=\bigg(\prod_{l<j}\prod_s\sigma_{s,l}^z\bigg)\bigg(\prod_{s'=1}^2\sigma_{s',j}^z\bigg)\sigma_{3,j}^-,\\\nonumber
		&\vdots\\
		c_{p,j}&=\bigg(\prod_{l<j}\prod_s\sigma_{s,l}^z\bigg)\bigg(\prod_{s'=1}^{p-1}\sigma_{s',j}^z\bigg)\sigma_{p,j}^-.
	\end{align}
\end{subequations}
Crucially, in contrast to the single-species Jordan--Wigner transformation, an \textit{on-site} string is incorporated in addition to its off-string counterpart. In our ED implementation, we have two matter sites ($N=2$). The $\mathrm{U}(2)$ QLM includes $p=6$ different species of fermions (two for the matter field, and two for each rishon). We therefore build a basis out of $pN=12$ Pauli spins, and the fermionic operators are constructed according to Eqs.~\eqref{eq:JW}.

\section{Results for different initial states and model parameters}\label{app:DiffInitState}
As mentioned in the main text, our conclusions for both the full and single-body protection seem to be independent of the particular choice of the initial state and hold for generic values of the parameters $\mu/J$ and $h/J$. Here, we provide ED results in support of this.

\begin{figure}[htp]
	\centering
	\includegraphics[width=.75\textwidth]{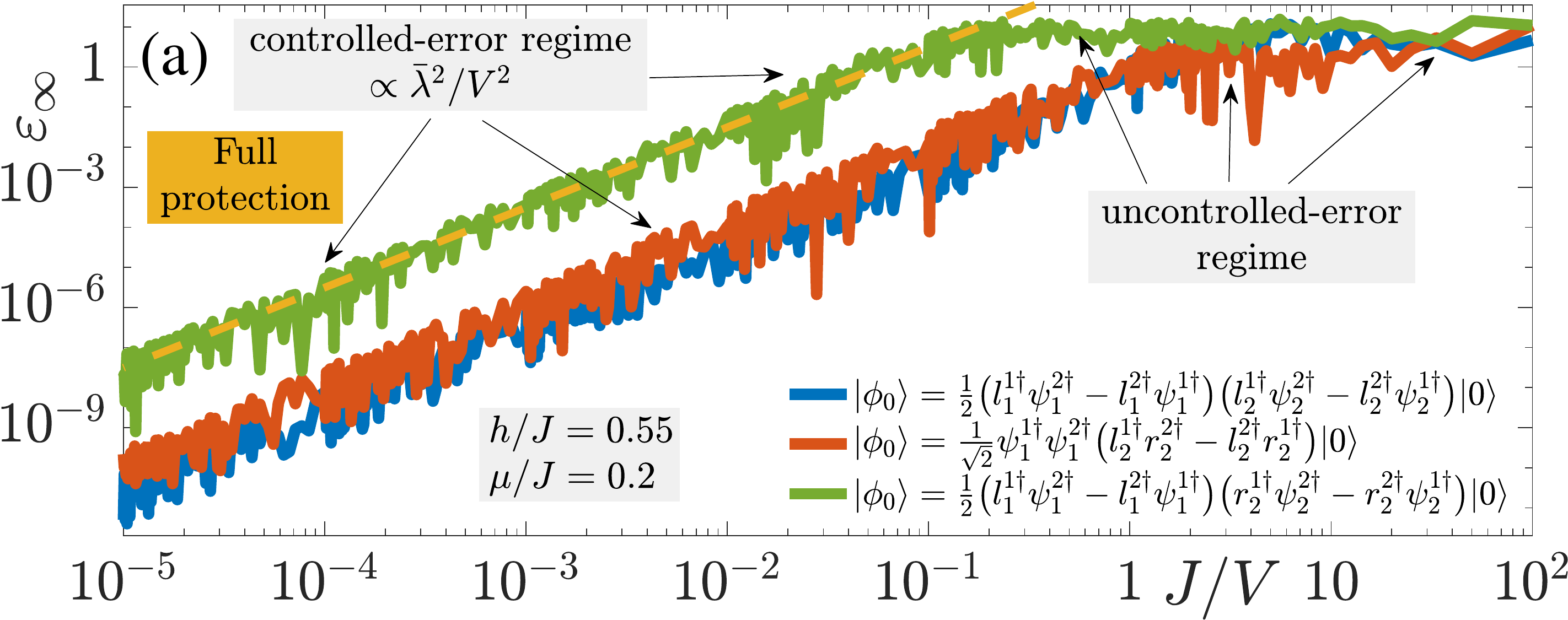}\\
	\vspace{1.5mm}
	\includegraphics[width=.75\textwidth]{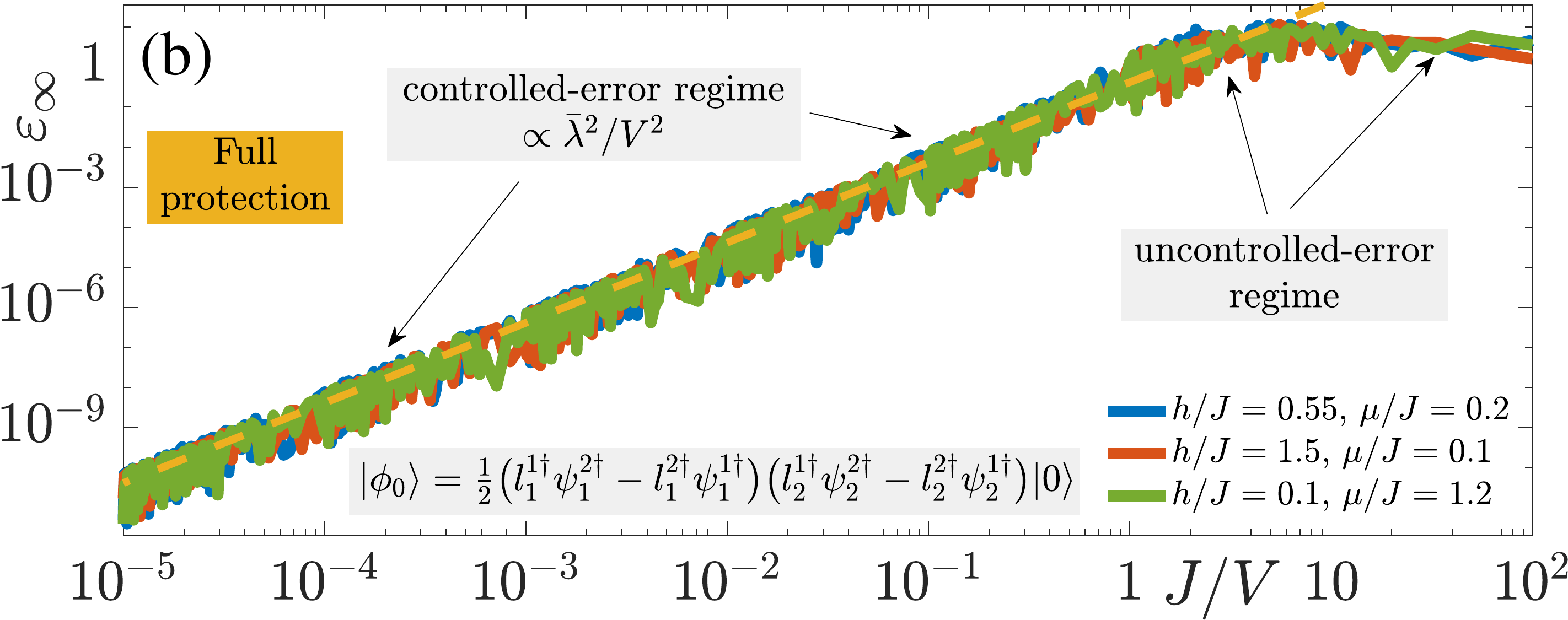}
	\caption{(Color online). Independence of qualitative conclusions on initial state and values of microscopic parameters. (a) Quenches of different initial states $\ket{\phi_0}$ (see legend) with the faulty theory $H=H_0+H_1+VH_G$, where $H_1$ is the error Hamiltonian whose terms are outlined in Table~\ref{table} and $VH_G$ is the full gauge protection given in Eq.~\eqref{eq:fullpro}. In all cases, we find a two-regime behavior for the infinite-time violation, such that it lies in an uncontrolled-error regime at sufficiently small $V$, whereas it enters a controlled-error regime when $V$ is sufficiently large. (b) The same independence is seen regarding the parameters $h/J$ and $\mu/J$ of $H_0$.}
	\label{fig:appFP} 
\end{figure}

\begin{figure}[htp]
	\centering
	\includegraphics[width=.75\textwidth]{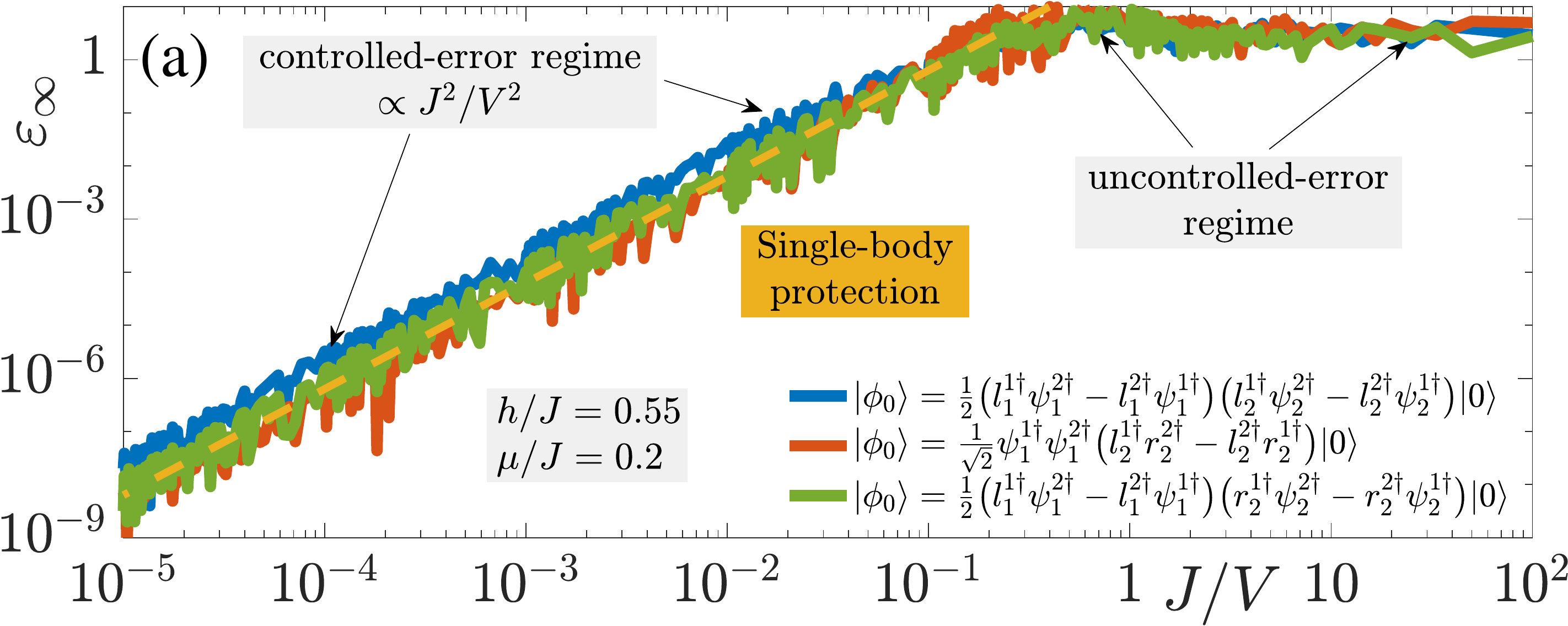}\\
	\vspace{1.5mm}
	\includegraphics[width=.75\textwidth]{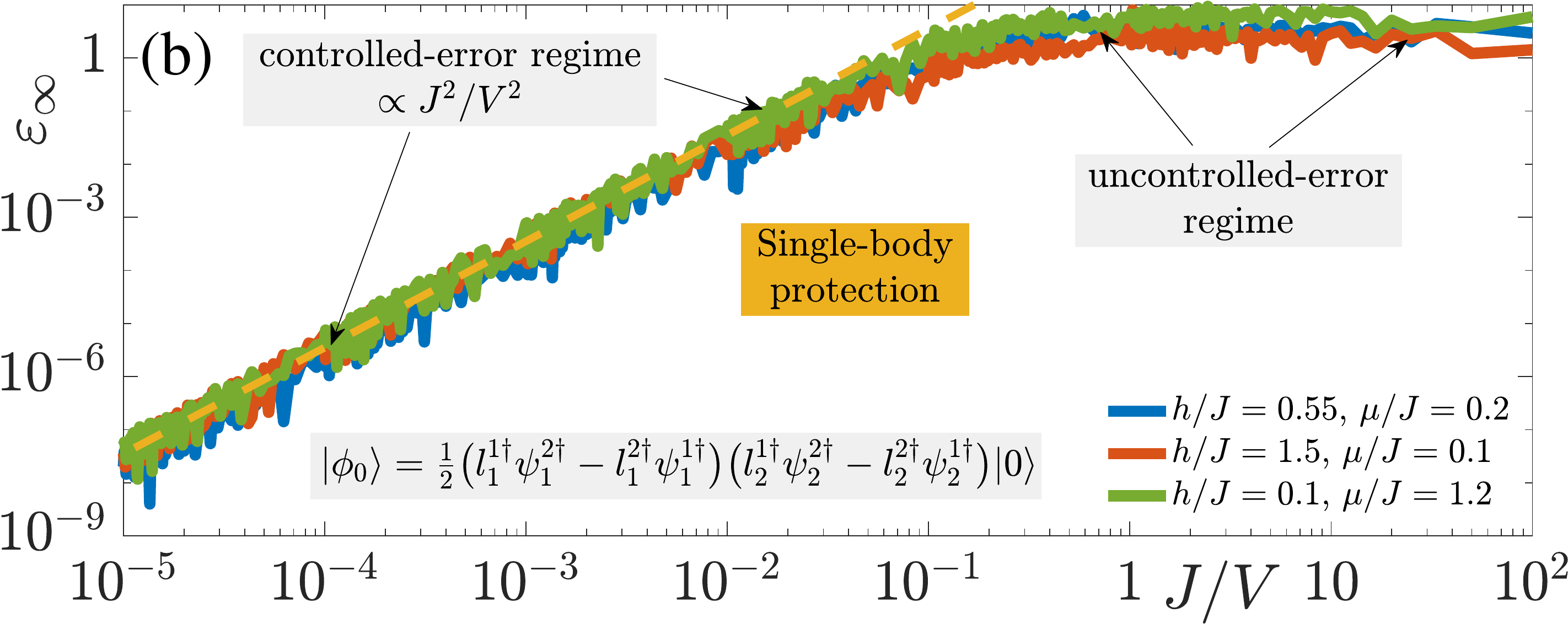}
	\caption{(Color online). Independence of qualitative conclusions on initial state. (a) Quenching any initial state $\ket{\phi_0}$ (see legend) with $H=H_0+\tilde{H}_1+V\tilde{H}_G$, where $\tilde{H}_1$ is the color-changing gauge-breaking term given in Eq.~\eqref{eq:ColorChanging} and $V\tilde{H}_G$ is the single-body protection defined in Eq.~\eqref{eq:LinPro}, we always find a two-regime behavior for the infinite-time violation, such that it lies in an uncontrolled-error regime at sufficiently small $V$, whereas it enters a controlled-error regime when $V$ is sufficiently large. (b) The same independence is seen regarding generic values of the parameters $h/J$ and $\mu/J$ of $H_0$.}
	\label{fig:appLP} 
\end{figure}

The infinite-time gauge violation is shown in Fig.~\ref{fig:appFP}(a) in the case of full gauge protection for three different initial states, one of which has been used for the results in the main text, given in Eq.~\eqref{eq:InitState}. Quenches of these initial states with the faulty theory $H=H_0+H_1+VH_G$, where $H_1$ is the error term whose components are outlined in Table~\ref{table} and $VH_G$ is the full protection defined in Eq.~\eqref{eq:fullpro}, lead to the same qualitative behavior in the infinite-time violation as a function of $J/V$. Independently of the initial state, the same picture of two regimes persists. At sufficiently small $V$, the infinite-time error is uncontrolled, i.e., it cannot be analytically connected to the protection strength. At sufficiently large $V$, the infinite-time error is controlled in that it scales $\sim\bar{\lambda}^2/V^2$, and can thus be analytically related to the protection strength. The same qualitative picture is attained in Fig.~\ref{fig:appFP}(b) when starting in the initial state of Eq.~\eqref{eq:InitState} and quenching with $H$ at different values of $h/J$ and $\mu/J$.

The results for the single-body gauge protection defined in Eq.~\eqref{eq:LinPro} also show robustness to initial conditions and different values of $h/J$ and $\mu/J$ in case of the gauge-breaking error of Eq.~\eqref{eq:ColorChanging}, as shown in Fig.~\ref{fig:appLP}.

Even though we predict analytically \cite{vandamme2021reliability} that at sufficiently large $V$ the full gauge protection will work in general independently of the nature of the unitary gauge-breaking error $H_1$, the initial state, or values of the parameters $\mu/J$ and $h/J$, we cannot guarantee the same for the single-body protection in the case of non-Abelian LGT. In the case of Abelian LGT, single-body protection can be made to work for any type of unitary error \cite{Halimeh2020e}, but the noncommutativity of the local generators renders single-body protection not as well-behaved in the non-Abelian case.

\section{Time-dependent perturbation theory}\label{app:TDPT}
We shall derive here the short-time scaling of the gauge violation using time-dependent perturbation theory. We denote the initial state in its density-matrix form $\rho_0=\ket{\psi_0}\bra{\psi_0}$, and begin our derivation with the von Neumann equation for time evolution
\begin{align}\label{eq:vonNeumann}
\dot{\rho}(t)=-i\big[H_0+\bar{\lambda} \bar{H}_1,\rho(t)\big]\equiv-i\big(\mathcal{S}_0+\bar{\lambda}\mathcal{S}_1\big)\rho(t),
\end{align}
where $\rho(t)=e^{-iHt}\ket{\phi_0}\bra{\phi_0}e^{iHt}$ is the density matrix of the system at evolution time $t$, and $\bar{\lambda}$ is the average of the error strengths of the terms comprising $H_1$, such that $H_1=\bar{\lambda}\bar{H}_1$ (see Table~\ref{table}). Furthermore, we have defined
\begin{subequations}
\begin{align}
\mathcal{S}_0\rho&\equiv\big[H_0,\rho\big],\\
\mathcal{S}_1\rho&\equiv\big[\bar{H}_1,\rho\big].
\end{align}
\end{subequations}
As such, one can write the formal solution to Eq.~\eqref{eq:vonNeumann} as
\begin{align}
\rho(t)=e^{-i(\mathcal{S}_0+\bar{\lambda}\mathcal{S}_1)t}\rho_0.
\end{align}
The Taylor expansion of this solution reads
\begin{align}\nonumber
&\rho(t)=\sum_{n=0}^\infty\big(\mathcal{S}_0+\bar{\lambda}\mathcal{S}_1\big)^n\frac{(-it)^n}{n!}\rho_0\\\nonumber
&=\bigg\{1+\sum_{n=1}^\infty\bigg[\mathcal{S}_0^n+\bar{\lambda}\sum_{m=0}^{n-1}\mathcal{S}_0^m\mathcal{S}_1\mathcal{S}_0^{n-m-1}+\bar{\lambda}^2\sum_{m=1}^{n-1}\sum_{k=0}^{n-m-1}\mathcal{S}_0^{n-m-k-1}\mathcal{S}_1\mathcal{S}_0^k\mathcal{S}_1\mathcal{S}_0^{m-1}+\mathcal{O}(\bar{\lambda}^3)\bigg]\frac{(-it)^n}{n!}\bigg\}\rho_0\\
&\approx\bigg\{1-it\mathcal{S}_0-\frac{t^2}{2}\mathcal{S}_0^2-\bar{\lambda}\bigg[it\mathcal{S}_1+\frac{t^2}{2}\big(\mathcal{S}_0\mathcal{S}_1+\mathcal{S}_1\mathcal{S}_0\big)\bigg]-\bar{\lambda}^2\frac{t^2}{2}\mathcal{S}_1^2\bigg\}\rho_0.
\end{align}
Writing the gauge-violation operator as $\mathcal{G}=H_G/N$, we calculate $\Tr\big\{\mathcal{G}\rho(t)\big\}$ component by component up to second order:
\begin{subequations}\label{eq:TDPT}
\begin{align}
\Tr\big\{\mathcal{G}\rho_0\big\}=&\,0,\\\nonumber
-it\Tr\big\{\mathcal{G}\mathcal{S}_0\rho_0\big\}=&-it\Tr\big\{\mathcal{G}H_0\rho_0-\mathcal{G}\rho_0H_0\big\}\\
=&-it\Tr\big\{\big[\mathcal{G},H_0\big]\rho_0\big\}=0,\\
-\frac{t^2}{2}\Tr\big\{\mathcal{G}\mathcal{S}_0^2\rho_0\big\}=&-\frac{t^2}{2}\Tr\big\{\mathcal{G}H_0H_0\rho_0-2\mathcal{G}H_0\rho_0H_0+\mathcal{G}\rho_0H_0H_0\big\}=0,\\\nonumber
-i\bar{\lambda} t\big\{\mathcal{G}\mathcal{S}_1\rho_0\big\}=&-i\bar{\lambda} t\Tr\big\{\mathcal{G}\bar{H}_1\rho_0-\mathcal{G}\rho_0\bar{H}_1\big\}\\
=&-i\bar{\lambda} t\Tr\big\{\big[\rho_0,\mathcal{G}\big]\bar{H}_1\big\}=0,\\\nonumber
-\frac{\bar{\lambda} t^2}{2}\Tr\big\{\mathcal{G}\big(\mathcal{S}_0\mathcal{S}_1+\mathcal{S}_1\mathcal{S}_0\big)\rho_0\big\}=&-\frac{\bar{\lambda} t^2}{2}\Tr\big\{\mathcal{G}H_0\big[\bar{H}_1,\rho_0\big]+\mathcal{G}\bar{H}_1\big[H_0,\rho_0\big]\\
&+\big[\rho_0,H_0\big]\bar{H}_1\mathcal{G}+\big[\rho_0,\bar{H}_1\big]H_0\mathcal{G}\big\}=0,\\\nonumber
-\frac{\bar{\lambda}^2t^2}{2}\Tr\big\{\mathcal{G}\mathcal{S}_1^2\rho_0\big\}&=-\frac{\bar{\lambda}^2t^2}{2}\Tr\big\{\mathcal{G}\bar{H}_1\bar{H}_1\rho_0-2\mathcal{G}\bar{H}_1\rho_0\bar{H}_1+\mathcal{G}\rho_0\bar{H}_1\bar{H}_1]\big\}\\
&=\bar{\lambda}^2t^2\Tr\big\{\mathcal{G}\bar{H}_1\rho_0\bar{H}_1\big\}\neq0\,\,\,\text{(generically nonzero)},
\end{align}
\end{subequations}
where we have utilized the cyclic property of the trace, in addition to $\mathcal{G}\rho_0=\rho_0\mathcal{G}=0$ and $\big[\mathcal{G},H_0\big]=0$. All but the last component of Eqs.~\eqref{eq:TDPT} are zero. As such, the leading contribution to the gauge violation $\varepsilon=\Tr\big\{\mathcal{G}\rho(t)\big\}$ at short times is $\propto\bar{\lambda}^2t^2$, similarly to the case of Abelian lattice gauge theories \cite{Halimeh2020a}.

\section*{References}
\bibliography{nonAbelianBiblio}
\end{document}